\documentclass[twocolumn,showpacs,aps,prl,preprintnumbers,amsmath,amssymb]{revtex4}
\usepackage{graphicx}

\def\angst{\,{\rm \AA}}

\newcommand{\diff}{{\rm\,d}}

\newcommand{\arctanh}[1]{\operatorname{arctan}}
\newcommand{\ee}{{\rm e}}

\def\zno{ZnO($10\bar{1}0$)\,}

\begin{document}
\title{Electrostatic Field Driven Alignment  of Organic Oligomers on ZnO Surfaces}

\author{F. Della Sala}
\affiliation{National Nanotechnology Laboratory, Istituto Nanoscienze-CNR, Via per Arnesano, I-73100 Lecce, Italy}
\affiliation{Centre for Biomolecular Nanotechnologies, IIT,  Arnesano, Italy}
\affiliation{IRIS Adlershof,  Humboldt-Universit\"at zu Berlin, Newtonstrasse 15, 12489 Berlin, Germany}
\author{S. Blumstengel and F. Henneberger}
\affiliation{Institut f\"ur Physik, Humboldt-Universit\"at zu Berlin, Newtonstrasse 15, 12489 Berlin, Germany}

\date{\today}  
\def\abf{}
\begin{abstract}
We study the physisorption of organic oligomers on the ZnO($10\bar{1}0$) surface using first-principles density-functional 
theory and non-empirical embedding methods.
We find that both in-plane location and orientation of the molecules are completely determined by the 
coupling of their quadrupole moments to the periodic dipolar electric field present at the semiconductor surface. 
The adsorption is associated with the formation of a molecular dipole moment perpendicular to the surface,
 which bears an unexpected linear relation to the molecule-substrate interaction energy.
Long oligomers such as sexiphenyl become well-aligned with stabilization energies of several 100 meV 
along rows of  positive  electric field, in full agreement with recent experiments.  
These findings define a new route towards the realization of highly-ordered 
self-assembled arrays of oligomers/polymers on ZnO($10\bar{1}0$) and similar surfaces.
\end{abstract}

\pacs{Valid PACS appear here}

\maketitle

Hybrid structures made of conjugated organic molecules and inorganic semiconductors exhibit 
an enormous application potential as they combine the favorable features of both components 
in a single new material \cite{blumst08}. However, interfacing of organic molecules with the 
typically highly reactive semiconductor is a complex issue. Rupture and fragmentation are 
frequently observed leading to ill-defined interfaces \cite{lin02}. On the other hand, the 
electronic structure of the semiconductor surface might be exploited for developing novel 
strategies of molecular aggregation. In this Letter, we  demonstrate that the electrostatic 
interaction between the semiconductor and the $\pi$-electron system gives indeed rise to the 
self-assemblage of stable and highly ordered monolayers for a wide class of conjugated 
organic molecules.

The specific surface under consideration is the non-polar $(10\bar{1}0)$ crystal plane 
of ZnO. The chemistry of ZnO surfaces, see e.g. Ref. \onlinecite{woell07}, has been largely investigated in the 
context of catalysis \cite{catlow08} and, more 
recently, much attention is paid to the linkage with organic dyes and polymers, 
driven, e.g., by photovoltaic applications \cite{law05, dag08}. 
In particular, it has been found experimentally that p-sexiphenyl (6P) absorbs 
flat on the \zno surface with the long axis of the molecule perpendicular to the polar 
[0001] direction \cite{blumsteg10}. 
In this study, the hybrid interface has been formed entirely under ultra-high 
vacuum conditions suggesting that intrinsic features of the semiconductor-molecule system are behind 
that type of aggregation. The theoretical analysis presented below  not 
only confirms this conjecture but reveals systematic tendencies common to all oligomers
that can be used to engineer the growth of inorganic/organic structures.  

\begin{figure}[h]
\includegraphics[width=0.9\columnwidth]{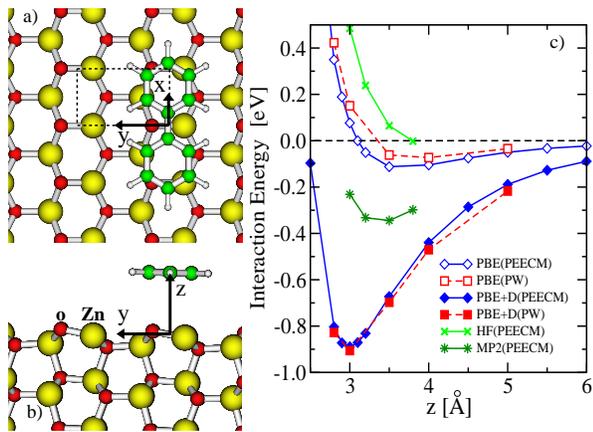}
\caption{a) Top and b) side view of 2P on the \zno surface (dotted rectangle: surface unit cell).     
c) Molecule-substrate interaction energy for 2P verus vertical distance z ($x$=$y$=0), as computed 
by different theoretical methods (see text).}
\label{fig1}
\end{figure}

In order to establish a proper and efficient methodical framework, we start 
with biphenyl (2P) as a short model oligomer. 
Fig. \ref{fig1}a and b depict the configuration examined. The origin of the 
reference coordinate system is located at the center of a surface Zn-O bond, the $z$- and $y$-axis 
point along the surface normal and the polar [0001] direction, respectively. 
The position of the molecule is denoted by the coordinates of its center of mass. 
We consider a clean, non reconstructed surface optimized using density functional theory (DFT) 
as described in Ref. \onlinecite{labat09}.
Our goal is the construction of the ground-state potential energy surface (PES) of 
the molecule-semiconductor system. 
In a first step, we set the center of the molecule 
on top of the Zn-O bond ($x$=$y$=0) with its long axis aligned in $x$-direction and the molecular plane parallel to the surface. 
The interaction energy of this arrangement keeping both the molecular and \zno surface configuration frozen 
is plotted versus distance from surface ($z$) in Fig. \ref{fig1}c. 
The curves are computed at the PBE level \cite{pbe} and with dispersion
correction (PBE+D) \cite{grimmedisp}, using two different computational methods: a periodic pseudopotential 
plane-wave (PW) approach\cite{suppmat} 
and the periodic electrostatic embedded cluster method (PEECM) \cite{PEECM,suppmat}, 
as implemented in the TURBOMOLE \cite{turbo} program.

Figure \ref{fig1}c shows that the PEECM results agree very well with the PW ones.
The practical advantage of PEECM lies 
in the fact that it considers only a single molecule in interaction with the surface. Unlike the PW method, where 
a whole periodic organic monolayer (of hypothetical structure) has to be treated, PEECM 
defines thus a cost effective way to tackle the initial adsorption step of the molecule. 
As expected  from previous studies, see e.g. Ref.  \onlinecite{rohlf}, the PBE functional leads 
to weak binding, while the 
dispersion correction increases the binding energy and reduces  the molecule-substrate distance. In order 
to verify the accuracy of PBE+D for ZnO surfaces, we performed 
reference MP2 calculations within the PEECM scheme\cite{suppmat}. 
Figure \ref{fig1}c exposes that PBE+D is quite far from MP2 and thus cannot be safely used for ZnO surfaces. 
The MP2 predicts  an interaction energy of 370 meV with an equilibrium distance $z_0\approx$ 3.5 $\angst$.
This molecule-substrate distance will be used in all the following calculations.

\begin{figure}[h]
\includegraphics[width=0.9\columnwidth]{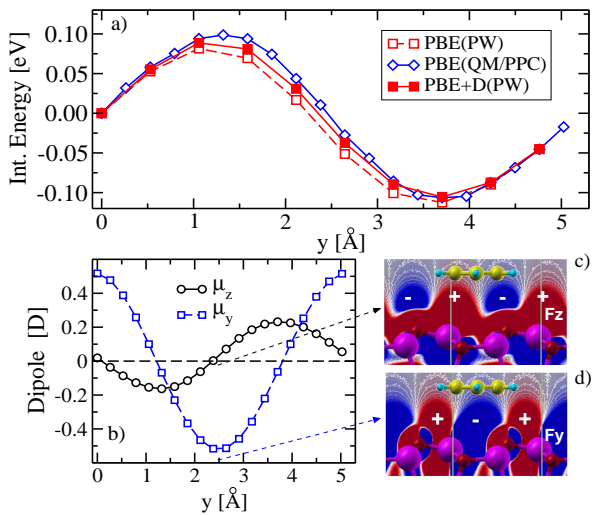}
\caption{
a) Molecule-substrate interaction energy versus $y$-position measured relative to $y=0$ $(x=0,z=z_0)$, 
by different theoretical methods. The length of the \zno surface unit cell in 
this direction is $5.19\angst$.  
b) Induced molecular dipole moment, components $\mu_y$ and $\mu_z$ ($\mu_x=0$) computed in QM/PPC. 
The electrostatic field of the \zno surface, from PW/PBE calculations, 
is illustrated (color online) in c) ($z$-component) and d) ($y$-component). 
The colormap covers the range from -5V/nm (intense blue) to 5V/nm (intense red).
Also shown is the molecule position at minimum $\mu_y$ ($y=2.5 \angst $). 
}
\label{fig2}
\end{figure}

As a next step, we now consider the change of the molecule-substrate interaction energy ($\Delta E$) when the 
molecule is translated along $y$-direction.
As displayed in Fig. \ref{fig2}a, the computations performed again in different approximations commonly 
reveal a distinct minimum for $y\approx$ 3.7\angst, i.e., when the center of the molecule is close to a position atop a Zn atom.
Interestingly, the relative energies of the PBE(PW) approach are practically  not modified by inclusion of the dispersion. 
In order  to save computation time for the examination of the complete PES below, we performed the same procedure 
but modeling the \zno surface only by {\it point-charges} with values $+q$ and $-q$ at the lattice positions of 
the Zn and O atoms, respectively. We call this method QM/PPC (quantum mechanics/periodic point-charges), because the 
molecule is treated quantum mechanically at  PBE level, while the ZnO surface is classically described. Hence,  
only the electrostatic interaction  between the molecule and the substrate is considered in this approach. 
For  $q$=1.2, excellent agreement with the PBE(PW) result is indeed achieved. This value $q$ 
is very close to what is  found in the Mulliken population analysis of the \zno surface \cite{labat09}. 
Therefore, we  conclude that exchange-correlation forces determine the {\it absolute} energy (see Fig. 1c), but the 
{\it  energy variation} when moving the molecule within the surface plane is completely dominated 
by the electrostatic coupling. Exchange-correlation effects vary on the atomic length scale, but are 
averaged out as the molecule is larger than the \zno unit cell. 

The alternating point charges which characterize the \zno surface create a {\it periodic dipolar  
electric field}  $\vec F$.  An important consequence of this field is that it generates in turn an {\it induced} dipole moment  
$\vec \mu$ in the 2P molecule. For symmetry reasons, $F_x$ is negligible, while $F_y$ and  $F_z$  reach 
values of  several V/nm. The resultant $\mu_y$ and $\mu_z$ are plotted versus $y$-position in Fig. \ref{fig2}b. 
Over the length $a$ of the unit cell, they change sign with a relative of shift of $a/4$. This behavior reflects 
the dipolar character of $\vec F$, as illustrated in Fig. \ref{fig2}c,d. 
The electric field is largely inhomogeneous, but sufficiently far from
the surface where the molecule is located, it has oscillating character.
When  $y \approx 2.5\angst$, $\mu_y$ reaches its negative maximum. At this position, as schematized 
in Fig. \ref{fig2}d, the molecule experiences a negative electric field over almost its whole size, 
whereas the average of $F_z$ is almost zero and thus $\mu_z \approx 0$.

\begin{figure}[h]
\includegraphics[width=0.8\columnwidth]{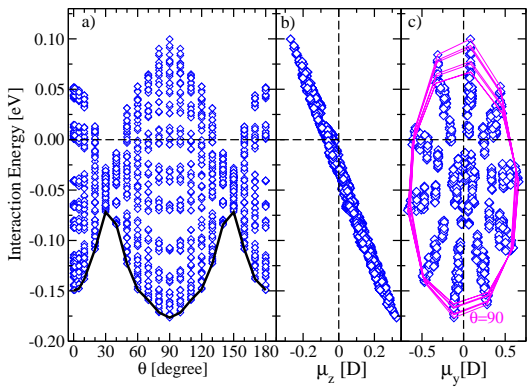}
\caption{Adsorption scenario of 2P on the \zno surface as computed by the QM/PPC method. 
a)  Molecule-substrate interaction energy versus rotation angle $\theta$ for all $x$- and $y$-positions counted.  
$\theta=0$: long molecular axis along $x$-direction (see Fig 1a).
In b) and c), the interaction energy is  plotted versus  the $z$- and $y$-component, respectively, of the induced dipole moment}
\label{fig3}
\end{figure}

The knowledge gained above enables us now to search for the global minimum of the PES by changing the residual 
degrees of freedom\cite{noteflat} - translation of the molecule along the $x$-direction and rotation around the $z$-axis.  
The QM/PPC approach makes it possible to scan a set of 1500 different molecular configurations. The results 
are condensed in Fig. \ref{fig3}.  
The QM/PPC interaction energy (relative to the isolated molecule) is represented in Fig. \ref{fig3}a 
as a function of the rotation angle $\theta$ for the whole set 
of $x$- and $y$-positions sampled over the \zno surface unit cell. 

The absolute minimum is found at $\theta=90^\circ$ (long 2P axis $\| y$).
However, there is also a second minimum at $\theta=0$, which corresponds to the one in Fig. \ref{fig2}a.
The energy difference between the two minima is only 20 meV and hence within the numerical error range. We conclude 
that the 2P molecule can be arranged on the \zno surface in two different cross-aligned orientations 
which makes the formation of a well ordered monolayer rather questionable.
 
The central question to be answered is about the mechanism controlling the alignment  of the molecule.  
The interplay between the surface electrostatic field, the induced dipole moment, and the interaction energy 
becomes evident from Fig. \ref{fig3}b and c. There is a distinct linear relation between the $\Delta E$ and $\mu_z$. 
The energy {\it is mimimized if and only if the dipole moment along $z$ is maximized}. 
On the other hand, as seen in Fig. \ref{fig3}c,  it holds $\mu_y\approx 0$ at 
the energy minimum, but there is no direct relation like in the case of $\mu_z$. 

The above findings become more transparent in an analytical model gathering the leading features 
of the molecule-substrate electrostatics. The energy of a molecule with zero static 
dipole but finite quadrupole moment $M_{ij}$  in a weak but {\it non-uniform} electric 
field ($F_x\approx 0$) is \cite{buck}
\begin{equation}
\Delta E \approx -\frac{1}{2} \sum_{i=y,z}  \left (
          M_{ii} \frac{\diff F_i}{\diff r_i}
 + \alpha_{ii}F_i^2   
 \right ),
\label{eq:enem}
\end{equation}
where  $\alpha_{ij}$ is the molecule's polarizability tensor. This expression is derived from 
perturbation theory: the first term represents the {\it electrostatic interaction} between 
the external non-uniform perturbing field and the unperturbed molecule, the second 
one the {\it induction energy} \cite{buck} accounting for
the molecular polarization created by the field.
Eq. (\ref{eq:enem}) is valid for a large class of planar, symmetric oligomers characterized by 
vanishing off-diagonal elements of $M_{ij}$ and $\alpha_{ij}$. 
The  \zno surface periodic dipolar electric field seen by the 
molecule can be approximated by (see Fig 2c,d and Ref. \onlinecite{suppmat})
\begin{equation}
F_y(y,z) \approx A \ee^{-k z} \cos (k y)  \,\, , \,\,
F_z(y,z) \approx -A \ee^{-k z} \sin (k y) \nonumber
\end{equation}
with $k=2\pi/a$ so that its norm ($F\approx A \ee^{-k z}$) is independent on $y$
and 
\begin{equation}
 \frac{\diff F_y}{\diff y} \approx k  F_z \;\;\;\  , \frac{\diff F_z}{\diff z} \approx - k  F_z.
\label{eq:grad}
\end{equation}
Using that $\mu_i=  \alpha_{ii} F_i $ and inserting (\ref{eq:grad}) in (\ref{eq:enem}), we obtain
\begin{equation}
\Delta E\approx - B \mu_z
 -C\mu_z^2   -\frac{\alpha_{yy}F^2}{2}
\label{eq:final}
\end{equation}   
with $B= k (M_{yy} - M_{zz})/{2\alpha_{zz}}$
and $C=(\alpha_{yy}- \alpha_{zz})/{2\alpha_{zz}^2}$.
\begin{table}[h]   
\begin{center}   
\begin{tabular}{l|rrrrcc}
\hline
\hline
molecule  & $M_{yy}$ & $M_{zz}$ & $\alpha_{xx}$ & $\alpha_{zz}$ &  B & C \\
           & a.u.      &   a.u.    &  a.u.        & a.u.            & eV/D & eV/D$^2$ \\
\hline 
2P      &  -46.2 &  -57.3 &  217.4  &  68.2 & 0.558 & 0.068 \\
6P      & -135.7 & -168.4 & 1257.6 & 190.5  & 0.589 & 0.062 \\ 
5A      &  -82.6 & -101.4 & 645.2  & 115.2  & 0.557 & 0.084 \\
5PV    & -145.3 & -180.1 & 2052.9 & 207.8  &  0.574 & 0.090 \\
\hline
\hline
\end{tabular}
\caption{\label{f_tab} PBE quadrupole moments ($M_{yy}, M_{zz}$), polarizabilities ($\alpha_{xx},\alpha_{zz}$) as well as 
B and C coefficients (see text) for different molecules.}
\end{center}
\end{table}

Table \ref{f_tab} compiles the values of the relevant parameters for 2P and three other representative 
molecules (6P, 5A=pentacene, 5PV=penta-phenylene-vinylene), all widely used in organic opto-electronics. 
Though the  anisotropy of $\alpha$ can be quite significant, 
the term quadratic in $\mu_z$  in (\ref{eq:final}), originating from the induction energy, is negligible 
against the linear term (i.e., the quadrupolar contribution) for weak fields. 
Hence, the analytical model fully recovers the numerical 
results of Fig. \ref{fig3}b. A linear fit to the data in  Fig. \ref{fig3}b for 2P provides a slope of  -0.48 eV/D in 
 very good agreement with the value expected from Table \ref{f_tab}. 

\begin{figure}[h]
\includegraphics[width=0.8\columnwidth]{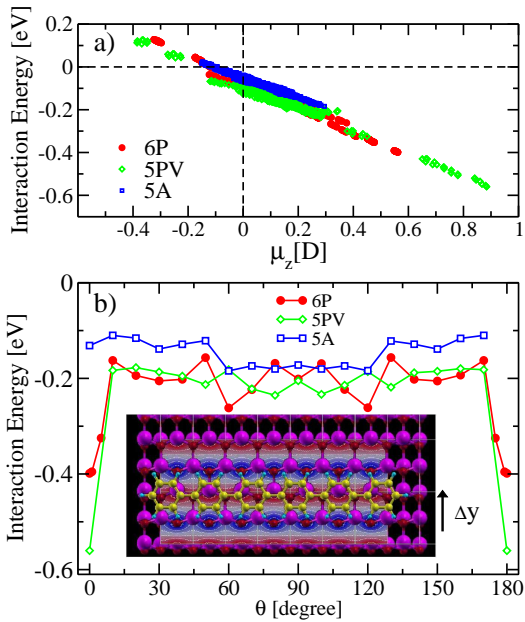}
\caption{Interaction energy for 6P, 5A and 5PV on \zno in QM/PPC.
a) Linear relation between molecule-substrate interaction energy and induced vertical dipole moment.
b) Interaction energy versus rotation angle, only the minimum values from all $x$- and $y$-positions scanned are shown. 
Inset (color-online): Orientation of 6P at the global minimum shown on a colormap
of the $z$-component of the surface electric field.
 }
\label{fig4}
\end{figure}

The analytical treatment suggests that the electrostatic scenario found for 2P is of general validity. 
Indeed, the full numerical analysis confirms this for 6P, 6A, and 5PV. 
A global optimization for molecules of this size cannot be carried
out using first-principles methods only and, usually, semi-empirical approximations have to be  
employed \cite{gulp}. The QM/PPC approach is instead non-empirical because the only parameter (q) is fixed 
from a first-principles (PBE/PW) calculations.
We thus were able to perform the same global scan of the PES as for 2P. Fig. \ref{fig4}a demonstrates 
that the linear relation between interaction energy and vertical dipole moment is a common 
feature for this class of non-polar molecules. Even, the slope of the curves is 
almost the same (-0.46 $\div$ -0.58 eV/D), consistent with the similar values of 
$B$ in Tab. \ref{f_tab}. 

Although the linear energy-dipole relation holds for all these molecules, the specific 
alignment on the substrate can be substantially different. This is documented in Fig. \ref{fig4}b, 
where the energy is plotted as a function of $\theta$ at $x$- and $y$-positions with minimized energy. 
In contrast to 2P, the PES of both 6P and 5PV exhibits deep global minima at $\theta=0$, clearly 
separated by 140 meV and 330 meV, respectively, from other arrangements. 
Not only the orientation but also the lateral position  of the adsorption site is uniquely defined, 
with $y \approx$ 3.7 $\angst$ and $x=0$ for both 5PV and 6P.
Thus, as illustrated in the inset for 6P, {\it the energy is minimized when the 
long axis of the molecule matches with the lines of largest positive $F_z$},  where the electrostatic coupling and  
thus $\mu_z$ are maximized. The longer the molecule, the more stable the alignment.
The PES of 5A is instead less deep and structured. In contrast to 6P/5PV, no preferred 
orientation can be thus anticipated here. This finding can be rationalized by 
the fact that 5A has {\it no carbon atoms exactly on the long molecular axis} which can be most easily polarized 
by the electric field, as it is for 6P/5PV.

In conclusion, we found that the periodic dipolar electric field of the \zno surface plays a key role 
in the adsorption of typical oligomers. 
When the molecules exhibit an axially oriented $\pi$-electron system, a well-defined 
molecular alignment{\abf , stabilized by energies  larger than 100 meV against reorientation, is established, 
as observed experimentally for 6P \cite{blumsteg10}.
The electrostatic coupling is characterized by a linear relation between the 
molecule-substrate interaction energy and the induced vertical molecular dipole moment, which
can be employed to predict and/or to design the molecular orientation on the surface.
Moreover, this dipole moment is directly associated with workfunction changes \cite{KRO07},
and thus provides a tool for engineering the energy level adjustment of inorganic/organic hybrid structures \cite{blumsteg10}.
Finally, we note that the single-molecule adsorption described above will be perpetuated and 
will result in molecular assemblies reflecting the topology of the surface field.
Although the induced dipole moment is modified by depolarization effects \cite{KRO07,spice}, this energy 
scale is certainly significantly smaller than the electrostatic molecule-substrate coupling controlling 
the alignment on the surface. Therefore, we believe that our findings define a route towards 
the realization of highly-ordered  self-assembled arrays of oligomer/polymers on \zno  and similar surfaces. }

\begin{acknowledgments}
We thank  R. Ahlrichs for providing us with the TURBOMOLE program package, M. Sierka, G. Heimel and I. Ciofini for discussions. 
This work is partially funded by the ERC Starting Grant FP7 Project DEDOM (no. 207441).

\end{acknowledgments}


\begin{thebibliography}{19}
\bibitem[{not()}]{noteflat}
\bibinfo{note}{Bending and inter-ring torsion or tilt with respect to the
  surface plane do certainly increase the total energy and can be thus safely
  ignored.}
\end{thebibliography}




\end{document}